\documentclass[review]{elsarticle}

\begin{document}

\begin{frontmatter}

\title{Segmented YSO scintillation detectors as a new ${\rm \beta}$-implant detection tool for decay spectroscopy in fragmentation facilities}

\author[utk]{R.~Yokoyama}
\author[utk]{M.~Singh}
\author[utk,ornl]{R.~Grzywacz}
\author[utk]{A.~Keeler}
\author[utk]{T.~T.~King}
\author[ific]{J.~Agramunt}
\author[ornl,utk]{N.~T.~Brewer}
\author[riken]{S.~Go}
\author[utk]{J.~Heideman}
\author[hku,riken]{J.~Liu}
\author[riken]{S.~Nishimura}
\author[proteus]{P.~Parkhurst}
\author[vnu,riken]{V.~H.~Phong}
\author[tntech]{M.~M.~Rajabali}
\author[ornl,utk]{B.~C.~Rasco}
\author[ornl]{K.~P.~Rykaczewski}
\author[ornl]{D.~W.~Stracener}
\author[ific]{J.~L.~Tain}
\author[ific]{A.~Tolosa-Delgado}
\author[agile]{K. Vaigneur}
\author[warsaw]{M.~Woli\'{n}ska-Cichocka}

	
\address[utk]{Department of Physics and Astronomy, University of Tennessee, Knoxville, TN 37996, USA}
\address[ornl]{Physics Division, Oak Ridge National Laboratory, Oak Ridge, TN 37830, USA}
\address[ific]{Instituto de Fisica Corpuscular (CSIC-Universitat de Valencia), E-46071 Valencia, Spain}
\address[riken]{RIKEN, Nishina Center, 2-1 Hirosawa, Wako, Saitama 351-0198, Japan}
\address[hku]{Department of Physics, the University of Hong Kong, Pokfulam Road, Hong Kong}
\address[proteus]{Proteus, Inc., Chagrin Falls, OH 44022, USA}
\address[vnu]{Faculty of Physics, VNU University of Science, 334 Nguyen Trai, Thanh Xuan, Hanoi, Vietnam}
\address[tntech]{Department of Physics, Tennessee Technological University, Cookeville, TN 38505, USA}
\address[agile]{Agile Technologies, Knoxville, TN 37932}
\address[warsaw]{Heavy Ion Laboratory, University of Warsaw, Warsaw PL-02-093, Poland}

\begin{abstract}
	A newly developed segmented YSO scintillator detector was implemented for the first time at
	the RI-beam Factory at RIKEN Nishina Center as an implantation-decay counter. 
	The results from the experiment demonstrate that the detector is a viable alternative
	to conventional
	silicon-strip detectors with its good timing resolution and high detection efficiency for
	${\rm \beta}$ particles.
	A Position-Sensitive Photo-Multiplier Tube (PSPMT) is coupled with a $48\times48$ segmented
	YSO crystal.
	To demonstrate its capabilities, a known short-lived isomer in $^{76}$Ni and the
	${\rm \beta}$ decay of $^{74}$Co were measured by implanting those ions into the YSO detector.
	The half-lives and ${\rm \gamma}$-rays observed in this work are consistent with the
	known values.
	The ${\rm \beta}$-ray detection efficiency is more than 80~\% for the decay of $^{74}$Co.
	
\end{abstract}


\end{frontmatter}


The study of ${\rm \beta}$-decays far from stability is essential to understand the
evolution of nuclear structure and nucleosynthesis processes.
It has been shown that, ${\rm \beta}$-decay properties of neutron-rich nuclei heavier than iron
impact $r$-process calculations that determine the solar abundance of elements
\cite{Mumpower2012,Surman2015}.
Typically, ${\rm \beta}$-decay experiments with such exotic nuclei involve intense cocktail beam
from fragmentation facilities.
The role of an implantation detector in these experiments is to measure the energy and the
positions of both heavy ion implantation and ${\rm \beta}$-ray emission in order to correlate
the identified ion with ${\rm \beta}$-decay	events.
Conventional implantation detectors developed for the beams at fragmentation facilities
use double-sided silicon-strip detectors (DSSDs) such as AIDA \cite{Griffin2014} or WAS3ABi
\cite{Nishimura2013} which were used recently with the BRIKEN
\cite{Tarifeno-Saldivia2017,Tolosa2019,Yokoyama2018} neutron counter array at the Radioactive
Ion Beam Factory (RIBF).

The idea of a new implantation detector using segmented scintillator crystals was proposed
previously \cite{Alshudifat2015}, aiming to use it as a fast ${\rm \beta}$ trigger for the
neutron	time-of-flight (ToF) array, VANDLE \cite{Peters2016}.
Energy information of ${\rm \beta}$-delayed neutrons can be deduced from neutron ToF
measurements \cite{Madurga2016}.
The DSSDs cannot be used for this application because of their slow response.
The new detector constructed using scintillator crystal is expected to enable neutron ToF
measurements at avanced fragmentation facilities that can provide high-intensity exotic beams.
A prototype plastic scintillator based detector described in Ref.~\cite{Alshudifat2015} was
used as the fast trigger for timing investigations at the NSCL \cite{Crider2016}.
Plastic detectors suffer from low stopping power for ${\rm \beta}$ particles which adversely
impacts the ability to make the ion-decay correlations.

In this paper, development of a new implantation detector using segmented YSO (Yttrium
Orthosilicate, Y$_2$SiO$_5$) crystals is presented.
There are several advantages of using YSO.
It offers better timing response than DSSDs and better
energy resolution compared to plastic scintillators. 
The effective atomic number ($Z=35$) and the density (4.4~g/cm$^{3}$) of the YSO crystal
are higher than silicon, resulting in higher stopping power, and therefore leading to shorter
range of ${\rm \beta}$ particles emitted by the ions implanted into the detector.
Due to these properties, YSO detectors are expected to have higher detection efficiency and
better position resolution for high-energy ${\rm \beta}$ rays than silicon-strip detectors.
Another advantage of using YSO scintillators is that they can be used with higher
implantation rates since they are radiation hard and the fast scintillation
(42~ns decay time \cite{Melcher1996}) enables rapid re-triggering.
The YSO is non-hygroscopic, relatively cheap and easy to make into pixelated arrays.
The design presented here is very easy to implement in an experiment. The use of resistive
readout requires only five 
detector signals for processing. The smaller light yield for heavy ions is typical for inorganic
scintillators, as measured for example in Ref. \cite{Koba2015}, which reduces the required
dynamic range of the light sensor needed to record signals
induced both by ${\rm \beta}$ particles and high-Z implants.

The segmented YSO detector was implemented for the first time at the RIBF to measure
${\rm \beta}$ decays of nuclei around and beyond $^{78}$Ni.
Analysis and results on the isomer of $^{76}$Ni and the ${\rm \beta}$ decay of $^{74}$Co nuclei
are presented as a demonstration of the capability of this new implantation detector.
The ${\rm \beta}$~efficiency for the $^{74}$Co decay and ${\rm \gamma}$-ray absorption for the
$^{76}$Ni isomer are also shown. 

\begin{figure}[htb]
	\includegraphics[width=\columnwidth]{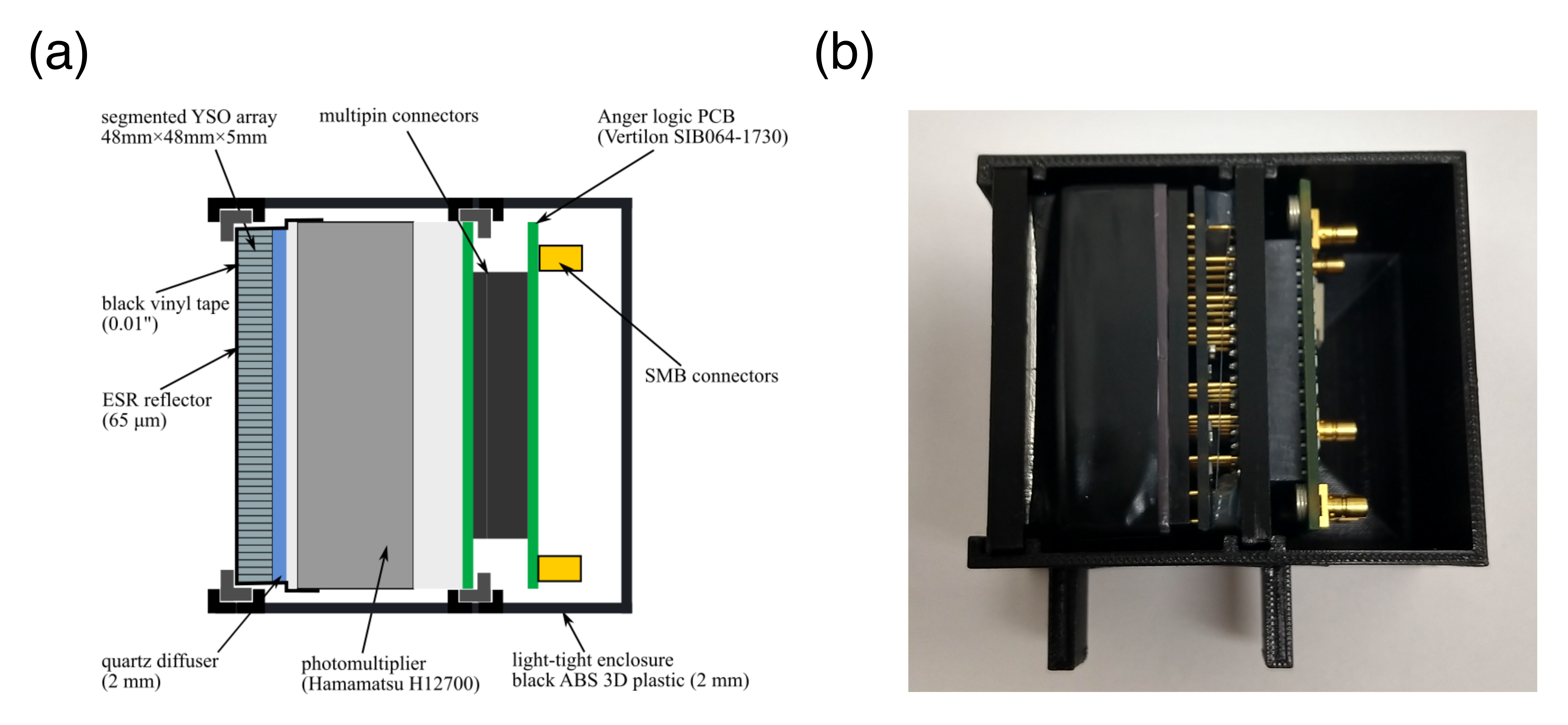}
	\caption{\label{schematic} 
		Schematic drawing (a) and a picture (b) of the YSO detector device.
		Detailed dimensions of the H12700 photomultiplier can be found on its data sheet
		\cite{H12700}. Schematic of the Anger-logic can be found in \cite{Vertilon}.
	}
\end{figure}

The detector unit consists of a segmented YSO crystal coupled to a flat panel multi-anode
photomultiplier by Hamamatsu, H12700.
The YSO scintillator is a high-granularity matrix of $1\times1$~mm segments arranged into a
$48\times48$ array in the $x$-$y$ plane.
The isolation of the pixel segments of the array using 65~${\rm \mu}$m ESR (3M) reflector
material enhances the position resolution, the sides, as well as the face of the crystal,
also have the ESR covering to prevent scintillation light loss.
The thickness of the crystal is 5~mm which is designed to stop all the ions of interest in the
$^{78}$Ni region produced by in-flight fission of $^{238}$U at the RIBF.
A 2~mm thick quartz light diffuser between the YSO crystal and the PSPMT was required in order to
achieve sub-anode position resolution by spreading the light from one pixel onto several anodes
of the PSPMT.
The detector unit was enclosed in a 3D-printed light-tight box with a thin light shielding black
vinyl tape covering the face of the detector.
The 64 anode channels of the photomultiplier are read out by a resistive-network, also known as
Anger-logic \cite{Anger1958}.
In this implementation, Vertilon SIB064-1730 (a custom version of the SIB064B-1018 \cite{Vertilon}) was
used as an Anger-logic read-out board.
The board is connecting the 64 anode outputs with a resistor network and is read out from its
four corners, which enables ${x}$- and ${y}$-position reconstruction from ratios of the
signal amplitudes.
The dynode signal is used to determine the time and the total energy.
Each of the four anode signals and the dynode are split into channels with two
different gain settings.
One set of signals is amplified 20 times by a high bandwidth amplifier for ${\rm \beta}$-ray
counting and the other is for heavy-ion implantations.


The YSO implantation detector was implemented together with the BRIKEN neutron counter system
\cite{Tarifeno-Saldivia2017,Tolosa2019} at RIKEN, RIBF in order to study ${\rm \beta}$ decays
of the nuclei around and beyond $^{78}$Ni.
The neutron-rich nuclei are produced by in-flight fission of a primary $^{238}$U$^{86+}$ beam
with an energy of 345~MeV/nucleon, induced at a 4 mm thick $^{9}$Be production target.
The typical intensity of the primary beam was $\sim60$~pnA during the run.
Fission fragments are separated and identified in the BigRIPS in-flight separator
\cite{Kubo2003} on an event-by-event basis by their proton numbers ($Z$) and the mass-to-charge
ratio ($A/Q$).
These quantities are obtained by the measurement of the $B \rho$, time of flight (TOF), and
energy loss ($\Delta E$) in BigRIPS.
A detailed explanation of the particle identification at the BigRIPS is found in
Ref.~\cite{Ohnishi2010,Fukuda2013}.
There were $2\times10^{5}$ $^{76}$Ni and $1.5\times10^{4}$ $^{74}$Co ions implanted into the YSO
detector in $\sim140$ hours of run.

The secondary beam was transported to the final focal plane and implanted into the YSO
detector.
Four layers of the double-sided silicon-strip detector (DSSD) WAS3ABi\cite{Nishimura2013}
were installed upstream of the YSO detector.
The DSSD consists of sixteen 3~mm wide strips in both the $x$ and $y$ directions.
The typical rate of the ion implantation in YSO during the run was $\sim60$~cps.

The YSO and WAS3ABi detectors are placed at the center of the BRIKEN high-density polyethylene
moderator \cite{Tarifeno-Saldivia2017}.
The BRIKEN detector was operated in {\it hybrid mode}, which is composed of 140 proportional
counters filled with $^{3}$He gas for neutron detection \cite{Tarifeno-Saldivia2017} and two
ORNL clover-type HPGe detectors \cite{Gross2000} for high-resolution
${\rm \gamma}$-ray detection.

\begin{figure*}[htb]
	\includegraphics[width=\textwidth]{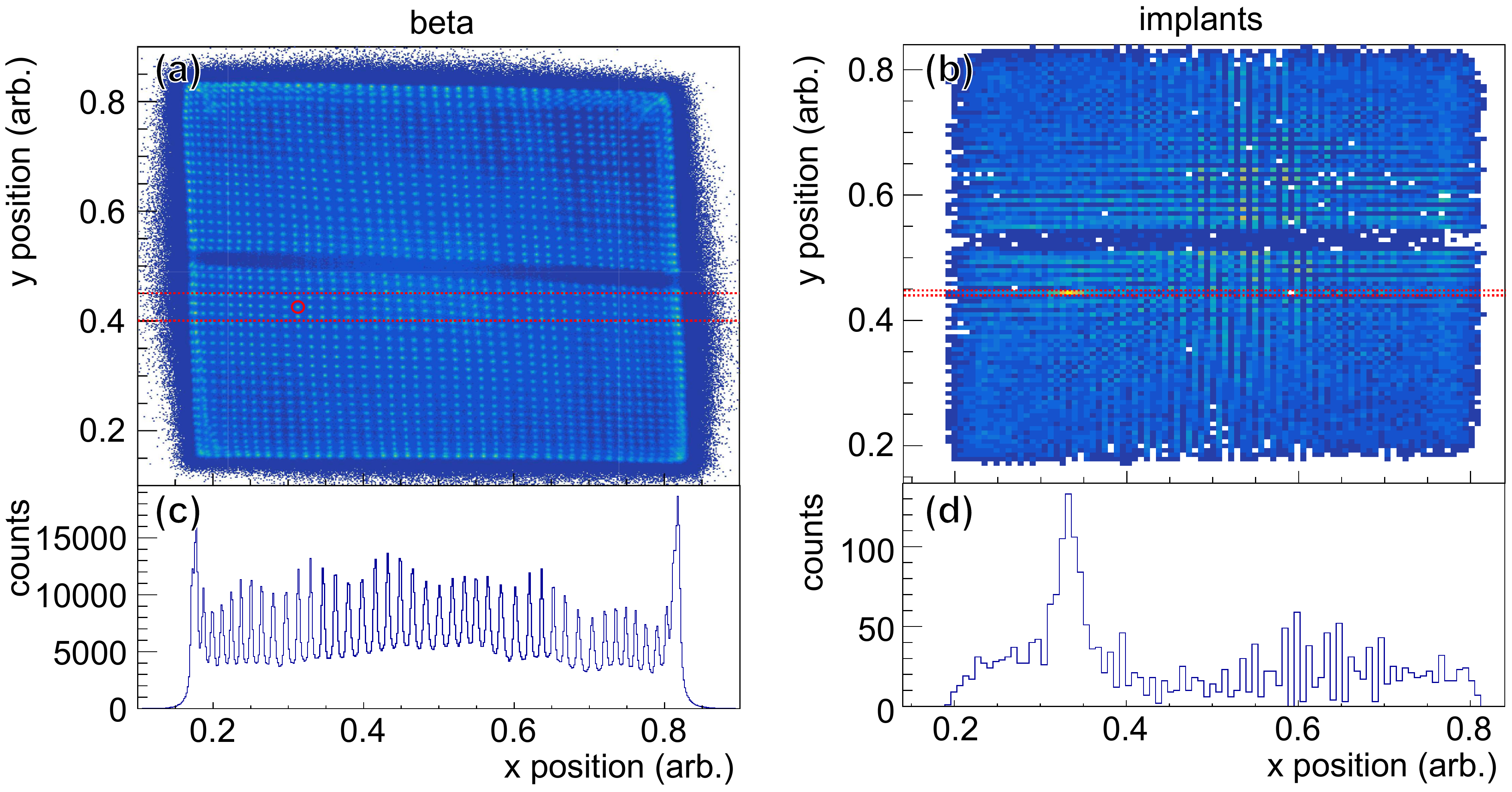}
	\caption{\label{position_image}
		YSO $x$-$y$ images of (a) ${\rm \beta}$ events and (b) implantation events correlated
		to the ${\rm \beta}$ events in a segment shown in the red circle in panel (a).
		(c) and (d) are the projection of (a) and (b), respectively, on to the $x$ axis in
		the cut shown by the red dashed lines. 
	}
\end{figure*}

As a result of the implementation of the YSO detector at the RIBF, we obtained position images for
${\rm \beta}$ and implantation events in the detector and successfully achieved the correlation
between recoil and decay events. 
The $x$-$y$ image for ${\rm \beta}$ events in the YSO detector is shown in 
Fig.~\ref{position_image} (a).
Each dot in the plot corresponds to a $1\times1$~mm segment of the YSO crystal.
The horizontal dip along $y\sim0.5$ is caused by the structure of the YSO segmentation.
The array is made of two parts of $24\times48$ segmented crystals and the ESR between those
two	in the middle distorts the image due to modification of light sharing between the two pieces.
The image has the resolution sufficient to distinguish segments for most of the active region.
Figure \ref{position_image} (b) shows the image of the implantation events occurring in the 1~s
before ${\rm \beta}$ events observed in one of the segments marked with a red circle in
Fig.~\ref{position_image} (a).
There is a clear peak on top of randomly correlated ion implantation events which are spread
over the entire detector surface.
The full-width half-maximum of the peak corresponding to the correlated implants shown in
Fig.~\ref{position_image} (d) is 1.3~mm.

\begin{figure}[htb]
	\includegraphics[width=\columnwidth]{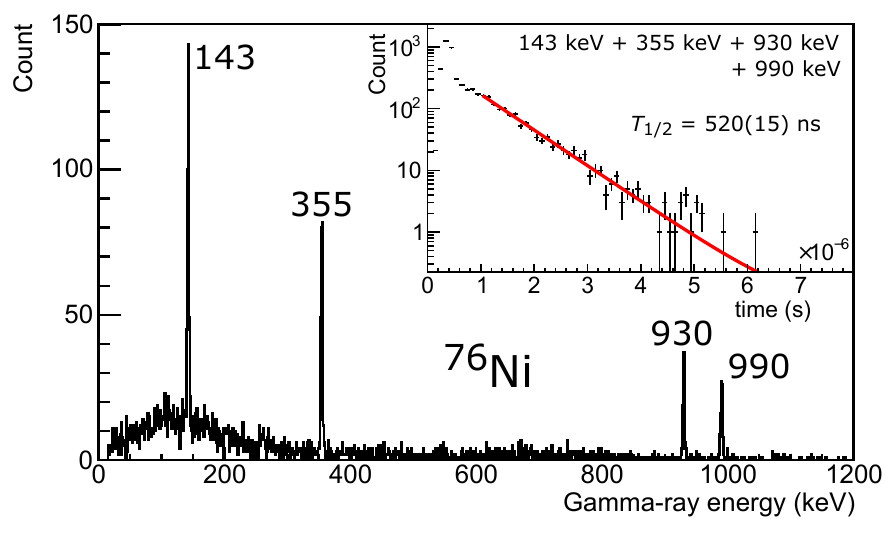}
	\caption{\label{isomer}
		Delayed ${\rm \gamma}$-ray spectrum of $^{76}$Ni gated from 1.2~${\rm \mu}$s to
		5~${\rm \mu}$s after the implantation.
		The time spectrum gated on the energy of all the four ${\rm \gamma}$-ray peaks is plotted
		at the right top corner of the figure.
		The red line shows the fitting function with an exponential decay and a constant
		background.
	}
\end{figure}

In $^{76}$Ni, an isomeric state is reported by Mazzocchi {\it et al.} produced by fragmentation
of a $^{86}$Kr beam in a $^{9}$Be target at NSCL \cite{Mazzocchi2005}.
Figure \ref{isomer} shows the delayed ${\rm \gamma}$-ray spectrum after $^{76}$Ni implantation
in the YSO detector.
The energy of the four ${\rm \gamma}$-ray peaks agrees with the literature \cite{Mazzocchi2005}.
The half-life of the isomer is obtained as 520(15) ns by fitting the decay curve of all the
four ${\rm \gamma}$-ray peaks, which is also consistent with the literature value,
$590^{+180}_{-110}$ ns \cite{Mazzocchi2005}.

Isomeric ${\rm \gamma}$-ray spectra of $^{76}$Ni implanted into YSO and WAS3ABi
detectors are compared in order to verify the effects of ${\rm \gamma}$-ray absorption in
each detector material. YSO is expected to attenuate ${\rm \gamma}$-rays more than silicon
due to its higher effective atomic number and density.
A simple simulation using Geant4 was performed to assess the severity of the absorption in YSO.
The results for the point source placed 6~cm apart from the surface of the clover detector
showed that the detection efficiency of 140 keV ${\rm \gamma}$-rays dropped by a factor of four
when the source is exactly placed at the center of the YSO crystal and when YSO is placed perpendicular
to the clover-detector surface.
However, in our experiment, we expected less absorption since the ion implantation is spread
over the crystal and there is effectively less material thickness than in the simulation.
For the case of the measured decay of $^{76}$Ni isomer, the ratio between the two peak areas,
that of 143~keV and 930~keV is (49(8))~/~(12(3)) = 4.3(15) for WAS3ABi,
whereas the ratio for YSO is (701(30))~/~(191(14))) = 3.7(3).
This shows that absorption in YSO detector of the 143 keV ${\rm \gamma}$-ray with respect to
the 930~keV line is only 1.2(4) times larger than that in WAS3ABi.

${\rm \beta}$-${\rm \gamma}$ spectrum of $^{74}$Co was first reported by Mazzocchi {\it et al.}
\cite{Mazzocchi2005}, then by Go {\it et al.} \cite{Go2018}, and recently by Morales {\it et al.}
\cite{Morales2018} with improved statistics.
Xu {\it et al.} \cite{Xu2014} and Hosmer {\it et al.} \cite{Hosmer2010} have also reported the half-life
of $^{74}$Co decay.
Hosmer {\it et al.} also measured $P_{n}$ branching ratio with BF$_{3}$ neutron counters
\cite{Hosmer2010}.

\begin{figure}[htb]
	\includegraphics[width=\columnwidth]{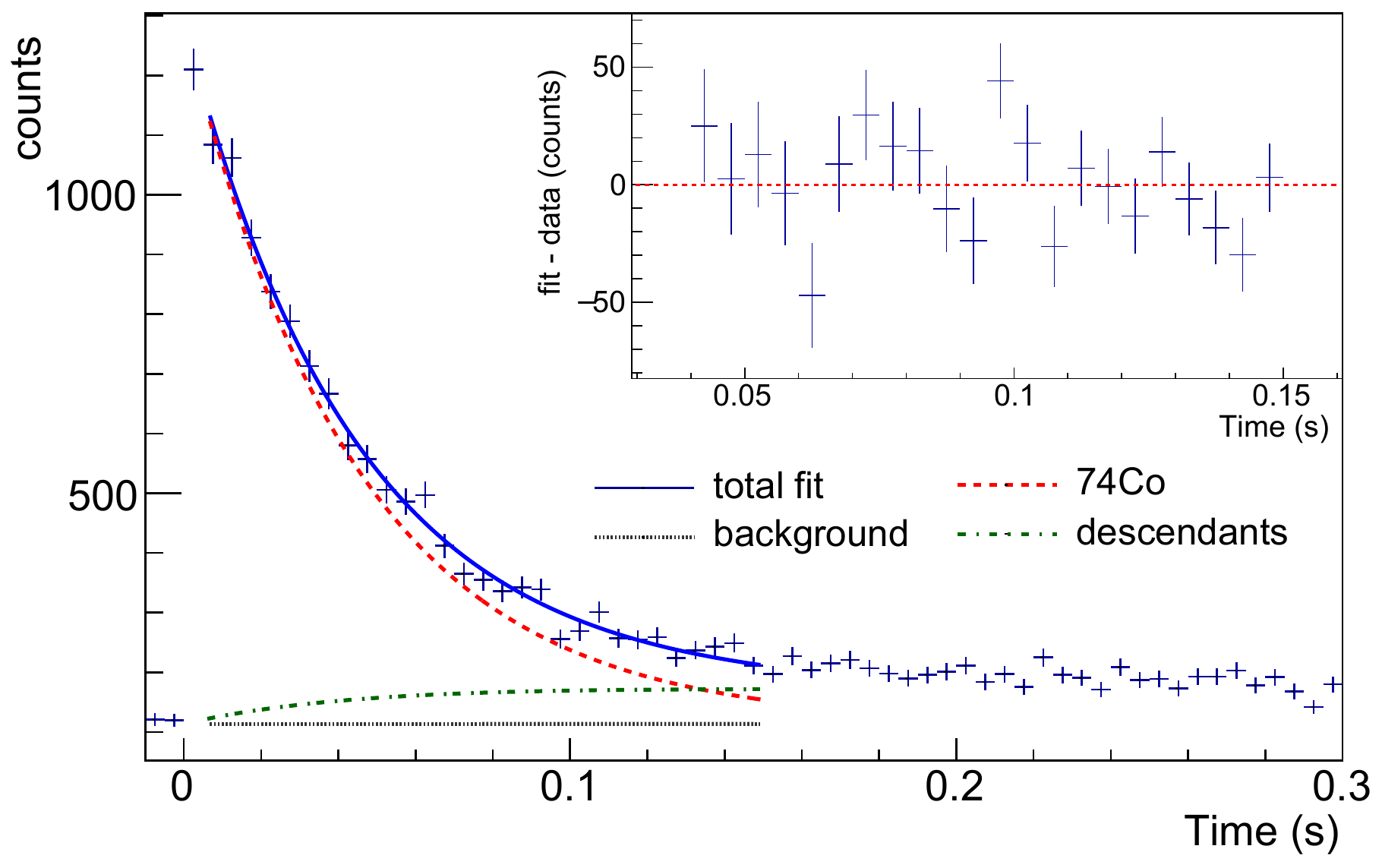}
	\caption{\label{decay}
		The decay curve of the ${\rm \beta}$ rays from $^{74}$Co decays. 
		The solid blue curve represents the fitting function.
		The dashed red curve shows the decay component of the parent nucleus, $^{74}$Co.
		The dashed-dotted green curve is the sum of the daughter and neutron-daughter branches.
		The dotted black line shows the linear background.
		The plot at the right top of the figure shows the difference between the data points
		and the fitting function.
	}
\end{figure}

Figure \ref{decay} shows the decay curve of $^{74}$Co ${\rm \beta}$ events.
The fit function includes the parent, daughter ($^{74}$Ni), grand-daughter ($^{74}$Cu), and
neutron-daughter ($^{73}$Ni) decays along with a linear background.
The half-lives of $^{74}$Ni, $^{74}$Cu, and $^{73}$Ni are fixed to the literature values,
0.5077~s \cite{Xu2014}, 1.59~s \cite{Winger1989}, and 0.84~s \cite{Franchoo2001}, respectively.
The $P_{n}$ branching ratio of $^{74}$Co is fixed to 18~\% \cite{Hosmer2010}.
The $^{74}$Co half-life obtained in this work is 30.8(6)~ms, which is consistent with the
literature values, 30(11)~ms \cite{Hosmer2010} and 31.6(15)~ms \cite{Xu2014}.
A ${\rm \beta}$-decay branch from isomeric state in $^{74}$Co is reported by Morales 
{\it et al.} \cite{Morales2018} with a 28(3)~ms half-life.

\begin{figure}[htb]
	\includegraphics[width=\columnwidth]{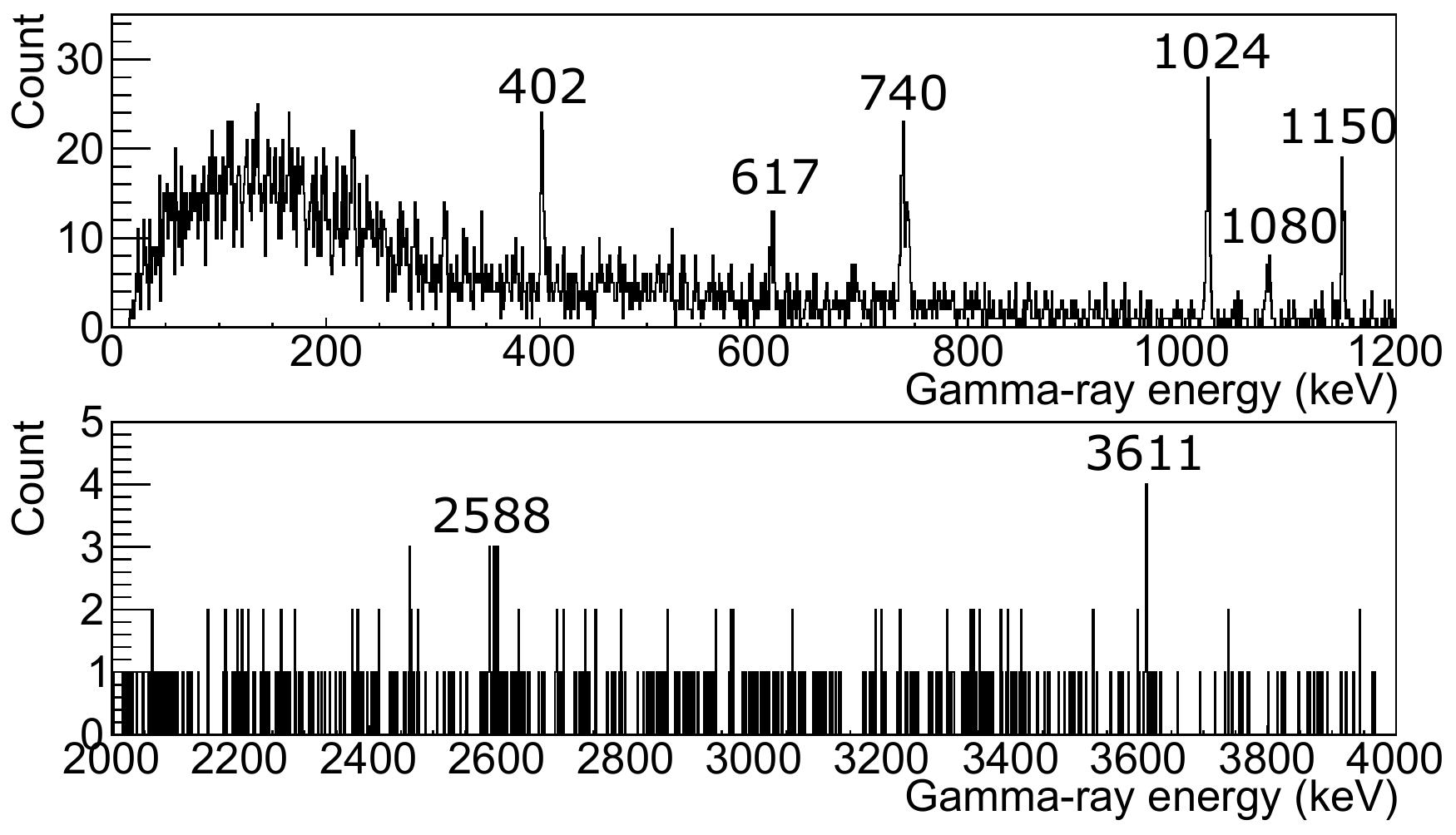}
	\caption{\label{decayEsp}
		The ${\rm \gamma}$-ray spectrum from the ${\rm \beta}$ decays of $^{74}$Co. 
		Energies of the peaks decaying by a similar half life to that of $^{74}$Co
		($\sim30$~ms) are labeled in keV.
	}
\end{figure}

\begin{figure}[htb]
	\includegraphics[width=\columnwidth]{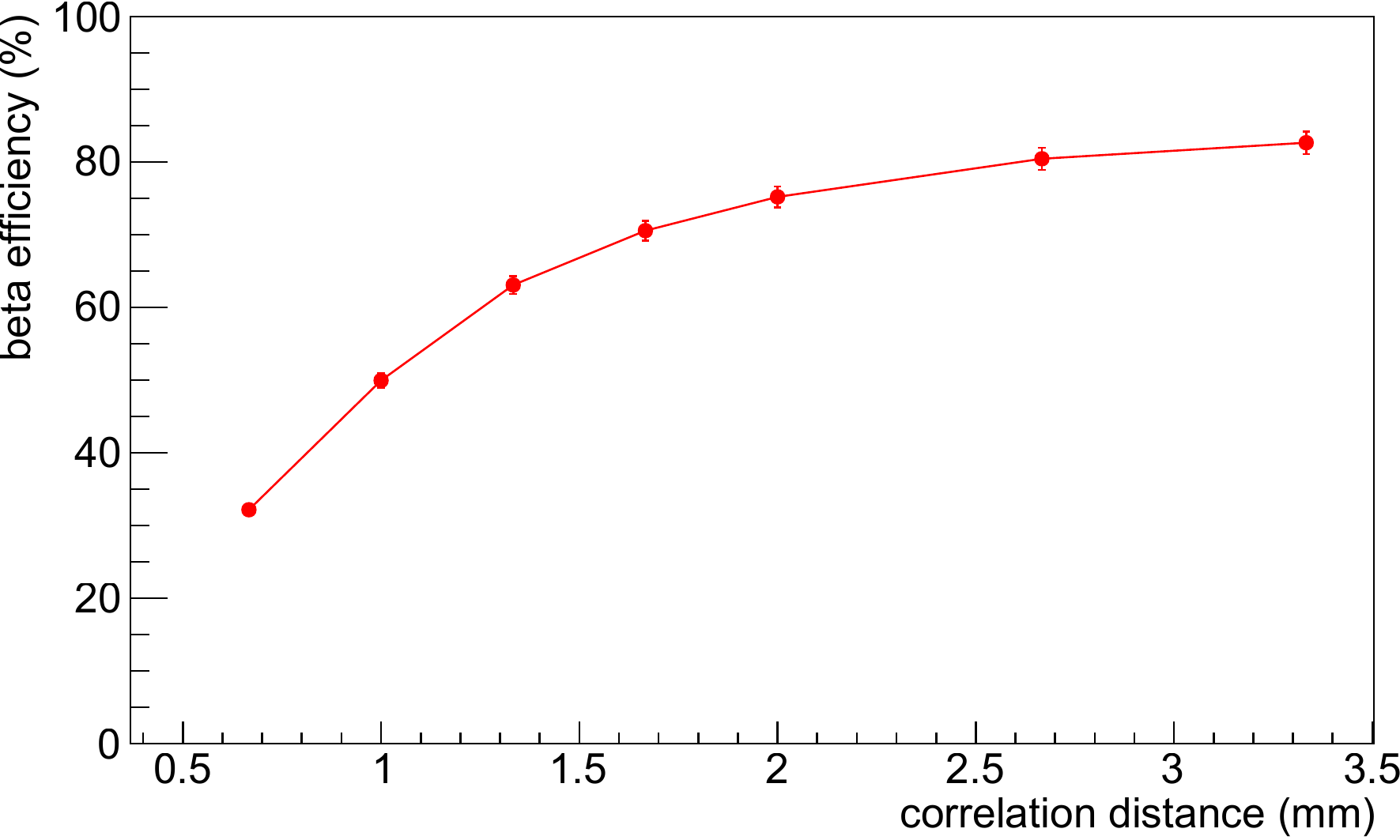}
	\caption{\label{beta_eff}
		The efficiency curve of the ${\rm \beta}$ rays from $^{74}$Co decays as a function of
		correlation	distances between ${\rm \beta}$ events and implant events. 
	}
\end{figure}

Figure \ref{decayEsp} shows the ${\rm \gamma}$-ray spectrum of $^{74}$Co decay detected within
100~ms after implantation.
All the ${\rm \gamma}$-rays in our spectrum are consistent with the reported values in
Ref.~\cite{Go2018,Morales2018}.
We also observed ${\rm \gamma}$-rays at 2588 and 3611~keV which are reported as decays from the
isomeric state in $^{74}$Co \cite{Go2018,Morales2018}.

The main purpose of using the segmented YSO detector in this experiment was to demonstrate
and exploit its high ${\rm \beta}$-detection efficiency.
Figure \ref{beta_eff} shows the ${\rm \beta}$-efficiency curve as a function of the maximum
correlation	distance between ${\rm \beta}$ events and implant events.
The correlation distance is calibrated to mm from the fact that the distance between two
adjacent segments is 1~mm.
The ${\rm \beta}$ efficiency is calculated by dividing the integral of the parent decay curve
obtained from fitting by the total number of $^{74}$Co implants.
The main source of the dead time is due to blocking ${\rm \beta}$ triggers for
200~${\rm \mu}$s after ion implantation in order to wait until the signal recovers from the
large pulse from heavy ions.
This corresponds to 1.2~\% dead time when the implantation rate is 60~cps.
The dead time of the data acquisition system during the run is neglected since it was much
smaller than 1~\%.
The ${\rm \beta}$ efficiency of the YSO detector for $^{74}$Co decay is more than 80~\% when
with maximum correlation distance grater than 3~mm, as shown in Fig.~\ref{beta_eff}.
This correlation range can be linked to the energy of the ${\rm \beta}$ particles in YSO.


In summary, we implemented the new segmented YSO detector for the first time in a
${\rm \beta}$-decay experiment at RIBF, RIKEN.
As a demonstration, data on the isomer in $^{76}$Ni and the ${\rm \beta}$ decay of $^{74}$Co
are shown.
The implant-${\rm \beta}$ correlation by the YSO detector was successful and obtained
30.8(6)~ms half-life for $^{74}$Co, which is consistent with previous reports 
\cite{Xu2014,Hosmer2010}.
We also confirmed the ${\gamma}$-rays of the $^{74}$Co decay reported in
Ref.~\cite{Go2018,Morales2018}.
The ${\rm \gamma}$-ray absorption in YSO is affecting the ${\rm \gamma}$-ray detection
efficiency, but not more than by a factor of 1.2(4) at 143~keV, compared to the
silicon-strip detector, WAS3ABi.
The ${\rm \beta}$-ray detection efficiency for $^{74}$Co is $\sim80\%$ with a 3~mm correlation
radius.
We have demonstrated that a compact and simple-to-operate segmented YSO detector can be
used as a very high-efficiency ${\rm \beta}$-implant and ion-${\rm \gamma}$ detection tool.
Also, a high stopping power equivalent to eight layers of 1-mm DSSDs makes it a much cheaper
option to use in ion-implantation experiments particularly when the range straggling of
implanted ions requires the use of a lot of active stopping material.
The segmented YSO crystal is in the same price range of a single DSSD detector.
The limited resolution of the scintillator does not impede ${\rm \beta}$-decay experiments.
In the situation when ions are fully stripped, the limited energy resolution for ion energy
measurement is also not a limitation.
YSO and other similar inorganic scintillators are known to be extremely radiation hard and
can be reused in many experiments.
The high stopping power for ${\rm \beta}$ particles resulted in very high detection efficiency.
Together with its excellent sub-nanosecond timing resolution, the YSO detector can be,
in many cases, a better alternative to DSSD detectors for ${\rm \beta}$-decay experiments at
fragmentation facilities.

\section*{acknowledgments}
The present experiment was carried out at the RI Beam Factory operated by RIKEN Nishina Center,
RIKEN and CNS, University of Tokyo.
This research was supported in part by the Office of Nuclear Physics, U.S. Department of Energy
under Award No.~DE-FG02-96ER40983 (UTK).

\section*{References}

\bibliography{MyCollection}

\end{document}